# Developments and the preliminary tests of Resistive GEMs manufactured by a screen printing technology

B. Clark[1], A. G. Agocs[2], R Oliveira[3], V. Peskov[3,4], Pietropaolo[3,5], Picchi[3,6]
[1]North Carolina State University, USA
[2] Eotvos University, Budapest, Hungary
[3] CERN, Geneva, Switzerland
[4] Ecole Superior des Mines in St. Etienne, France
[5]INFN Padova, Italy
[6]INFN Frascaty, Italy

## Abstract

We report promising initial results obtained with new resistive-electrode GEM (RETGEM) detectors manufactured, for the first time, using screen printing technology. These new detectors allow one to reach gas gains nearly as high as with ordinary GEM-like detectors with metallic electrodes; however, due to the high resistivity of its electrodes the RETGEM, in contrast to ordinary hole-type detectors, has the advantage of being fully spark protected. We discovered that RETGEMs can operate stably and at high gains in noble gases and in other badly quenched gases, such as mixtures of noble gases with air and in pure air; therefore, a wide range of practical applications, including dosimetry and detection of dangerous gases, is foreseeable. To promote a better understanding of RETGEM technology some comparative studies were completed with metallic-electrode thick GEMs.

A primary benefit of these new RETGEMs is that the screen printing technology is easily accessible to many research laboratories. This accessibility encourages the possibility to manufacture these GEM-like detectors with the electrode resistivity easily optimized for particular experimental or practical applications.

## I. Introduction

In the last decade, a great interest arose to various hole-type avalanche gaseous detectors: array of capillaries [1], capillary plates [2], gas electron multiplier (GEM) [3]. The unique feature characteristic of these novel detectors is their capability to operate at relatively high gains in poorly quenched gases and their capability to operate in cascade, which allows one to boost the overall gain. Capillary plates and GEMs offer considerable promise in many applications, such as constituting the basic element for gas photomultipliers [4, 5] in TPCs and as tracking devices [6].

For the last several years our focus has been centered on the development of a more robust version of the hole-type detector: Thick GEM (TGEM) [7, 8] and TGEM with oxide coatings [9]. The greatest success, however, was noticed when the electrodes of TGEM were manufactured from resistive materials. In the first attempt, a thick layer of graphite paint was used [10]; in the latest version we successfully tested resistive Kapton [11]. Both types of resistive detectors could operate at gains as high as TGEMs while offering the advantage of being fully spark-protected.

Unfortunately, it is not easy to obtain resistive Kapton from DuPont, which is the sole producer of Kapton. However, there is nothing magic with Kapton and certainly many other materials could be used to achieve the same spark-protective effect; this notion encouraged the development and testing of the first prototypes of RETGEMs made manufactured using screen printing technology.

Screen printing is widely used in microelectronics to produce patterns of different shape and resistivity. Therefore, RETGEM technology produced with screen printing techniques offers a convenient and widely available alternative to RETGEMs made of Kapton. This report serves to summarize the immediate results obtained in testing these new RETGEMs.

## II. Materials and methods

### II.1 The RETGEM manufacturing by a screen printing technology

A plate of DE-156, an Isola product often referred to as "G-10," was used as the base material. DE-156 is a halogen-free, glass epoxy laminate; it is coated on both sides with 17μm of copper (see Fig 1a).

The detectors were manufactured in the following consequent steps:

1) A photolithographic method was implemented to remove the excess copper from the top and bottom of the DE-156. The result was the creation of a copper border (see Fig 2b).
2) A resistive paste (Encre MINICO using for transistors printed circuits) is applied to the top and the bottom surfaces using screen printing technology. The paste is cured in air at 200° C for one hour. After the curing process is complete, the resistive layer was 15μm thick (Fig 1c).
3) Consistently sized holes were drilled at even intervals (using a CNC machine) in the region enclosed by the copper border (see Fig 1d).

In the work referred to in this paper, the detectors had the following geometrical and resistive characteristics: thickness was 1mm, active was area 30 x 30mm$^2$, hole diameter was 0.5mm, pitch was 0.8mm, resistive layer thickness was 15μm, the surface resistivity was 1 MΩ/□.

The photo of one of our detectors is presented in Fig. 2. Investigations by a microscope reveal a high quality of holes and boarders around them (see Figs 3 a and b).

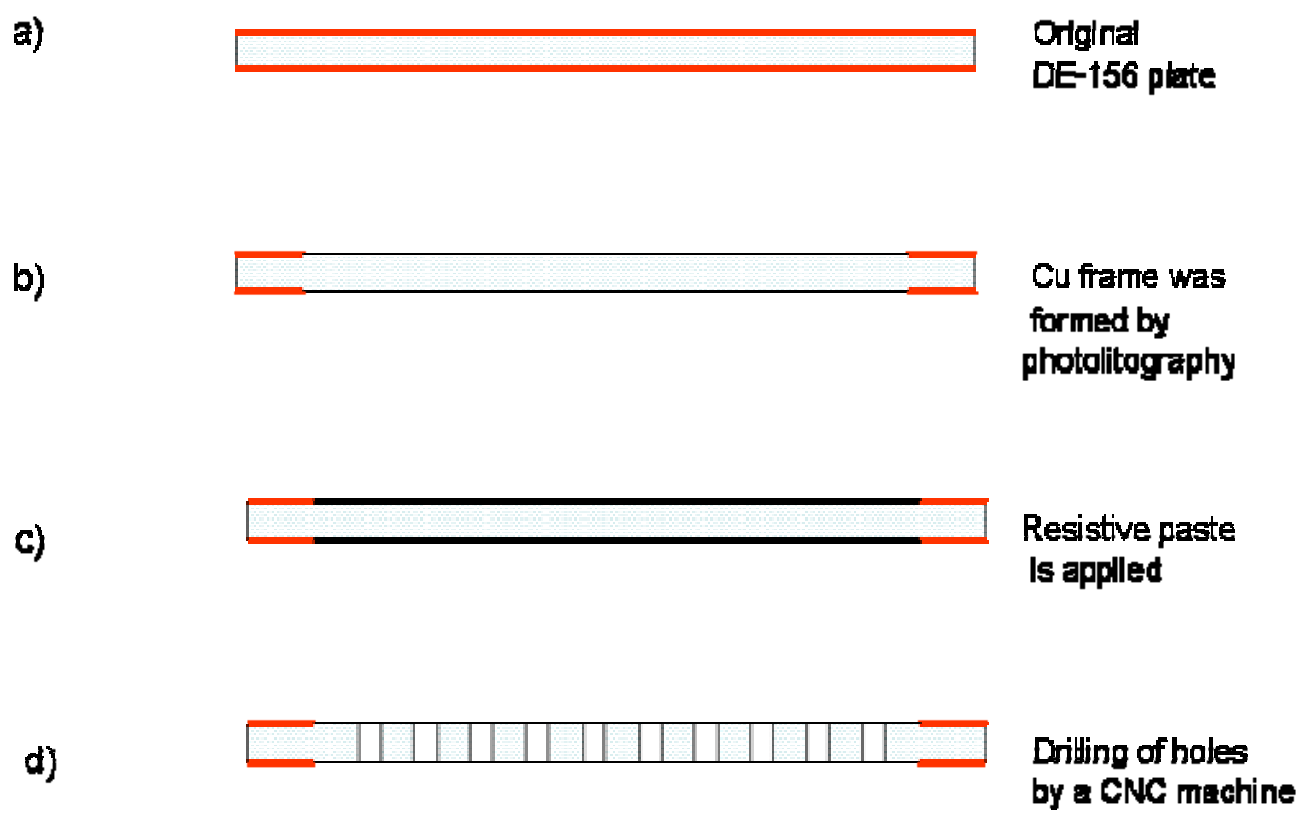

Fig. 1. Consequent steps in manufactring RETGEM using screen printing technoloy;
a) An original G-10 plae
b) Cu frame was manufatured
c) Resistive paste is applied
e) after heat treatment hole were drilled

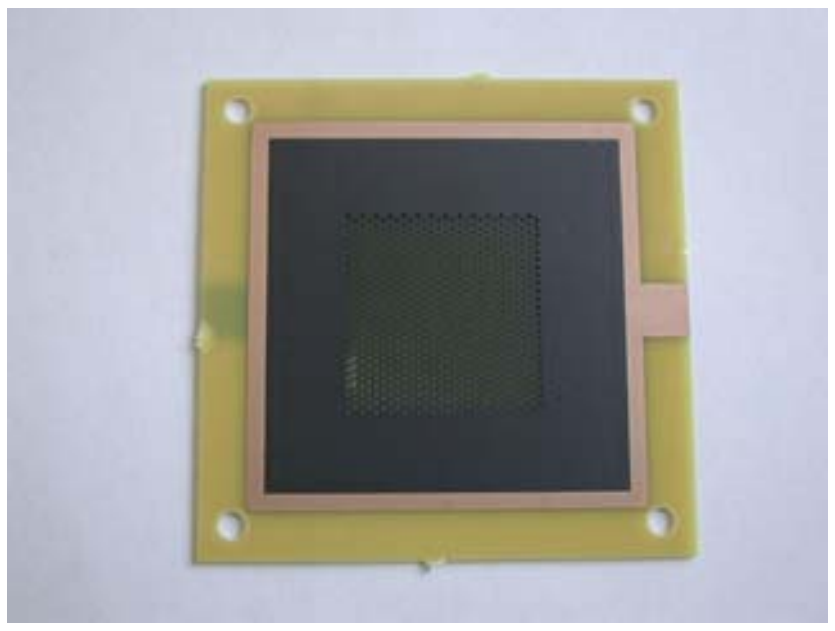

Fig. 2. A photo of the RETGEM produced using screen printing technology

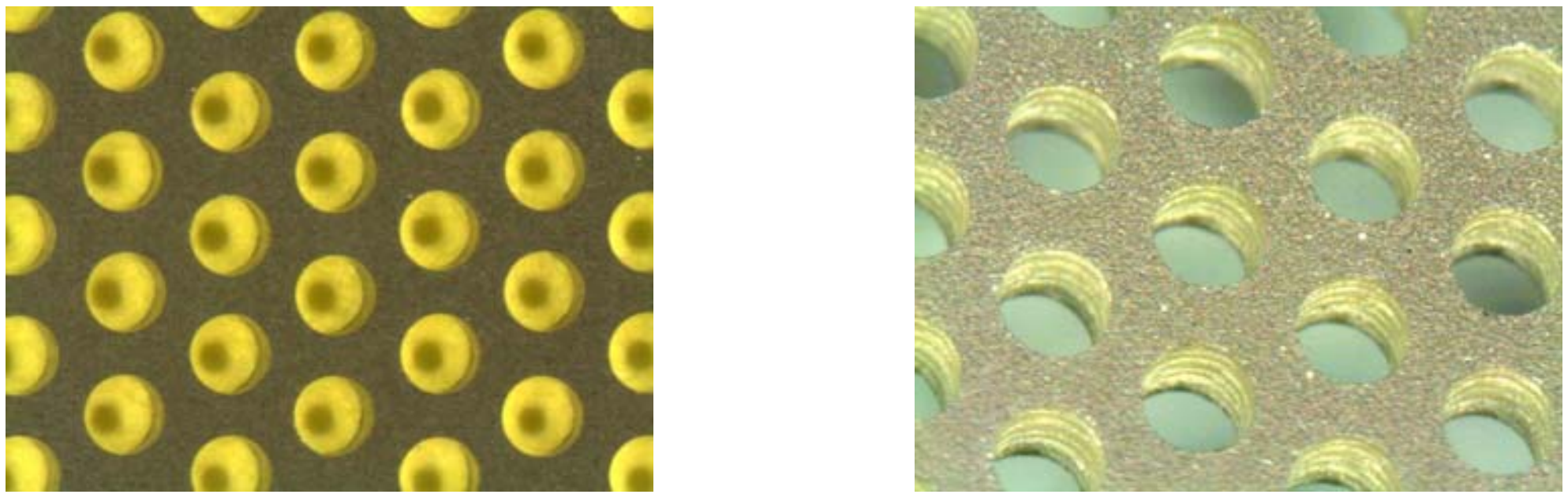

a) b)

Fig. 3. Photo of holes at various maginifications: a) medium magification, b) higher maginification

## II.2. The principle of operation

The principle of operation for this detector is the following: when a high voltage (HV) is applied to the Cu frames the resistive electrodes, due to their non-infinite resistivity, are charged up to a potential equal to that of the respective Cu frames and begin to act as equipotential layers (assuming the electrostatic case). Thus the same electric field is formed inside and outside of the holes as in the case of the TGEM with the metallic electrodes (see Fig .4). One can expect that at low counting rates the detector will operate as a conventional TGEM; whereas at high counting rates and in the case of discharges the detector's behavior will be more similar to that of the resistive plate chambers.

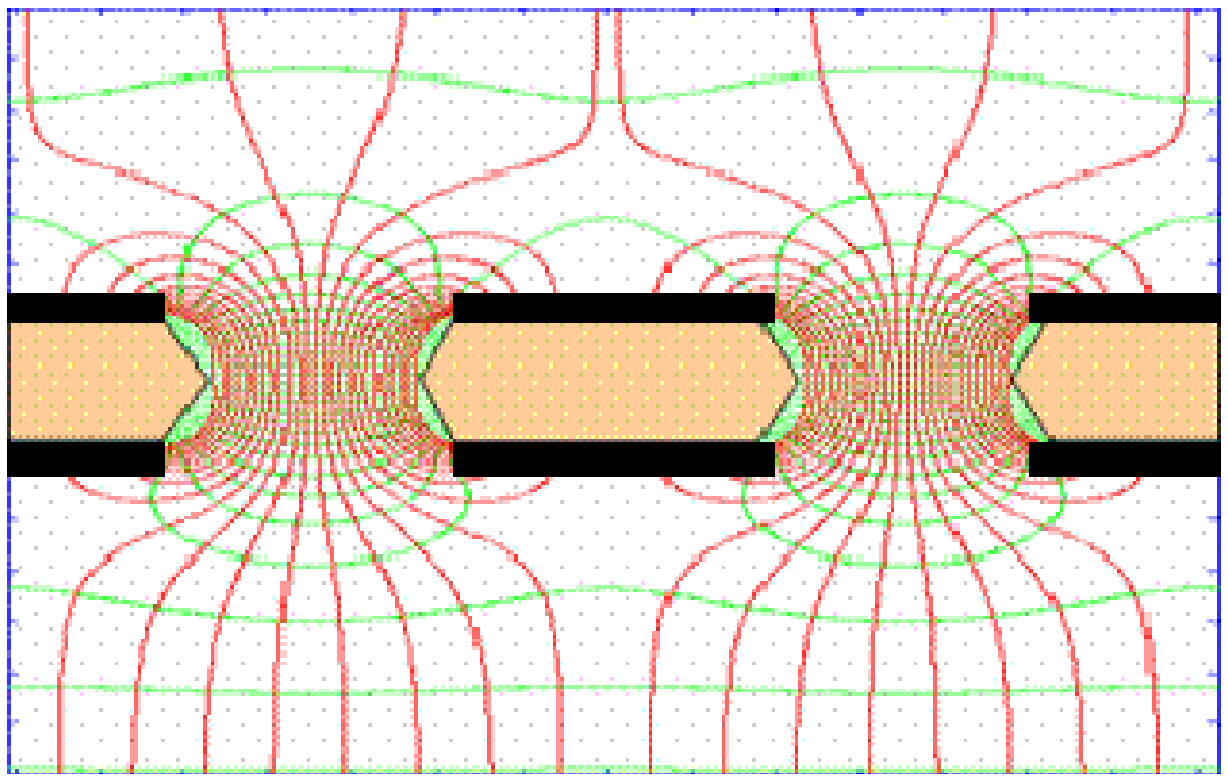

Fig. 4. Expected field line formation in the RETGEM

II.3. Experimental setup

Our experimental setup was essentially the same as described in [12] (see Fig. 5).

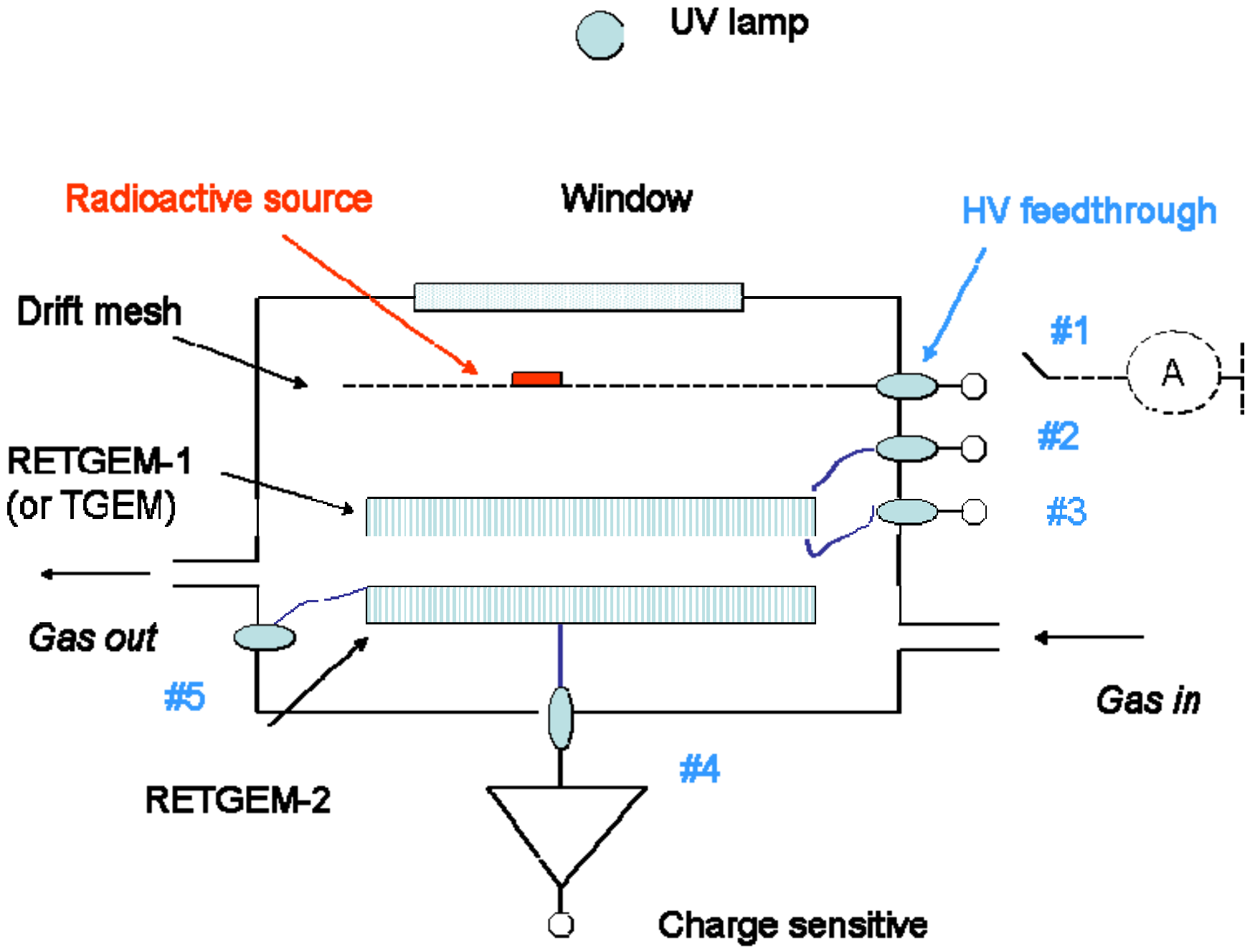

Fig. 5. A schematic drawing of the experimental se up

The setup consisted of a test chamber housing a RETGEM, or two RETGEMs operating in cascade, and a gas system, which allowed for the flushing of various gases through the chamber.  In the case of the double RETGEM setup, a resistive chain was implemented as shown schematically in Fig. 6.

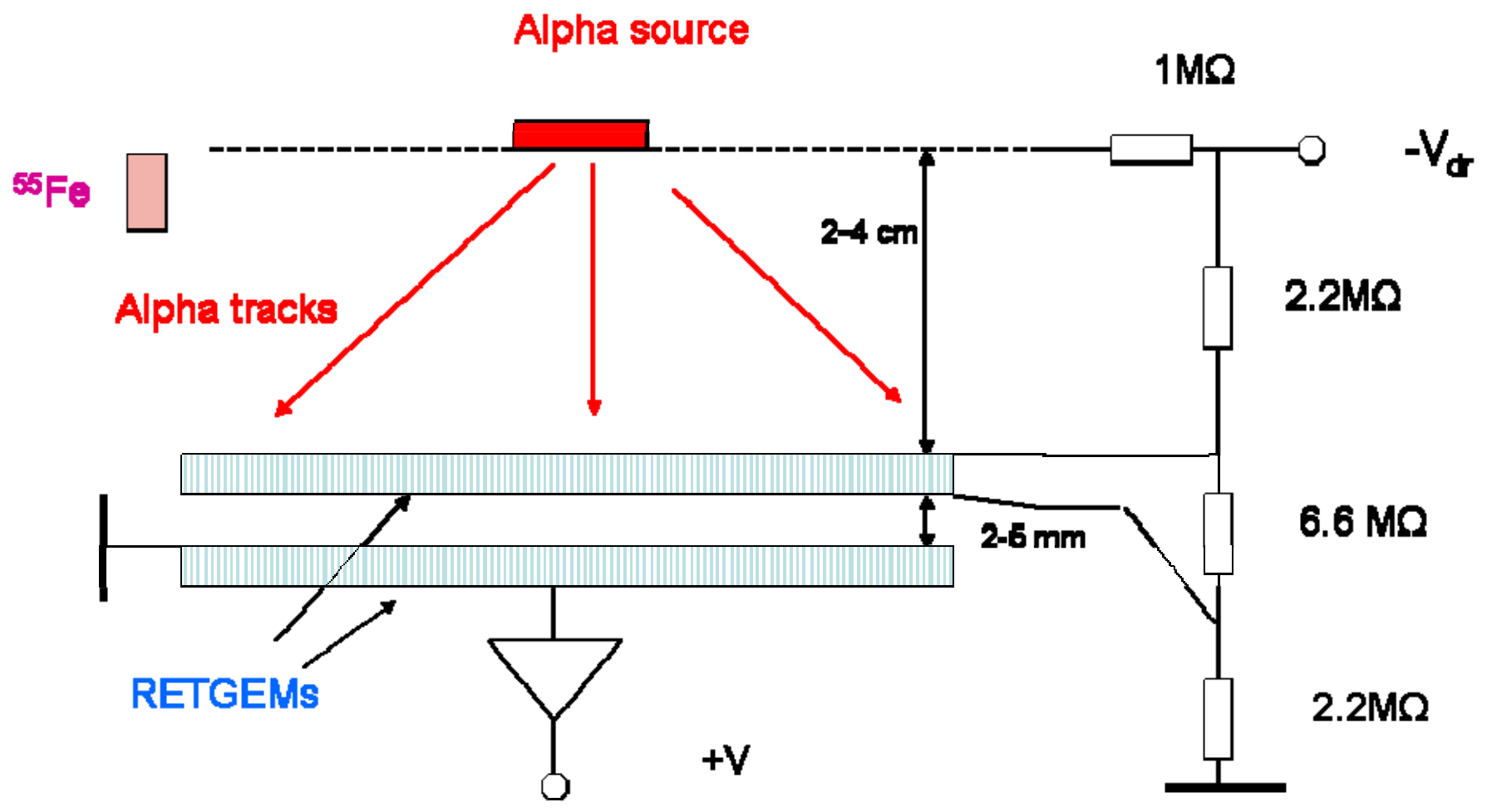

Fig. 6. A schematic drawing of the voltage divider used in noble gas, two-step RETGM measurements.

To insure a fair comparison of these new RETGEMs with Kapton RETGEMs, first tests were performed in the same gases as in the case of the Kapton RETGEM; these gases included Ne, Ar, and Ar+$CO_2$, which were all maintained at an approximate pressure of 1atm.

Some tests were completed in mixtures of Ar with air and in pure air. The latest studies involving air were motivated by possible applications of these new detectors in such practical areas as dosimetry, dangerous gas detection, and in a flame detection system. [13].

In some control measurements we also used metallic TGEMs having the same geometry as the RETGEMs referred to in this study.

Ionization of the gas was produced either by an $^{241}$Am alpha emitting source or by an $^{55}$Fe x-ray emitting source. The signals from the detector were recorded by a charge sensitive amplifier, Ortec 142PC or CANBERRA, and then, if necessary, additionally amplified by an Ortec research amplifier.

We also performed some measurements in current mode; a picoamperemeter, Kethley 487, was used for this purpose.

## III. Results

III.1 Tests in Ne, Ar and Ar+$CO_2$.

Fig.7 shows some results of measurements with the alpha source in Ne.

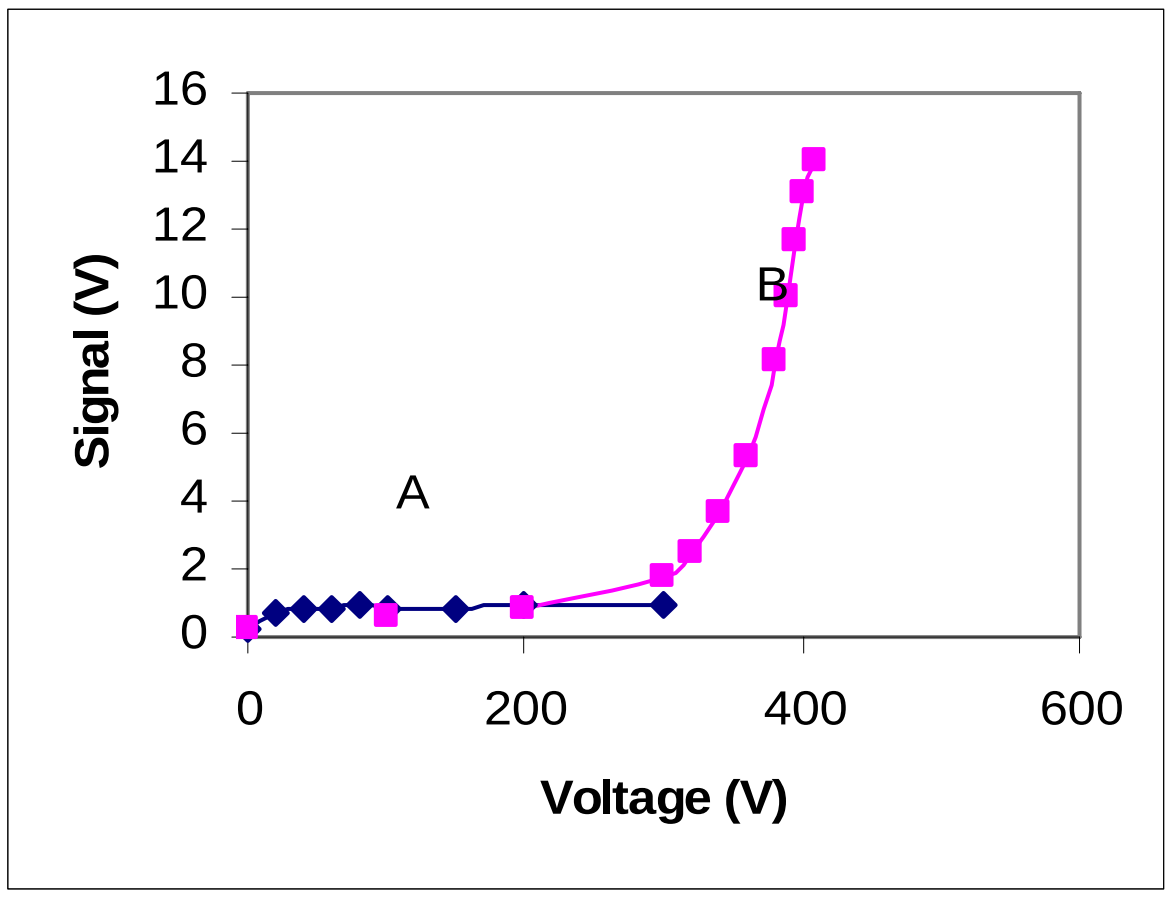

Fig.7. Results of measurements with alpha particles in Ne

Curve A represents the signal measured at feedthrough #2 (see Fig. 5) vs. the negative voltage applied to the feedthrough #1. One can see that at V> 100V the curve reached the saturated level (S=0.9 V) corresponding to full collection of charges, which were created by the incoming alpha particle, on the top electrode of the RETGEM-1. (Note that the RETGEM-# scheme simply refers to a specific detector used in the measurement. In a two-step setup, RETGEM-1 will refer to the top RETGEM and RETGEM-2 will refer to the bottom RETGEM.) Curve B corresponds to measurements of the signal amplitude vs. positive voltage applied to the bottom of the RETGEM-1 when the charge sensitive amplifier was connected to the feedthrough #3. During these measurements the voltage applied to the feedthrough #1 was -300V and the top electrode of the RETGEM was grounded via the feedthrough #2. One can see that at V>300V avalanche multiplication started in the holes of the RETGEM-1. The RETGEM gain A can be calculated as:

$$A=G/S, (1)$$

where G is the signal amplitude measures at feedthrough #3. Note that both our charges-sensitive amplifiers were calibrated by the charge injection method (see [9]) and the formula (1) allows us independently verify this calibration.

Fig. 8 shows the gain curves measured in Ne with 5.49MeV alphas ($^{241}$Am source) and 5.9 keV x-rays ($^{55}$Fe source).

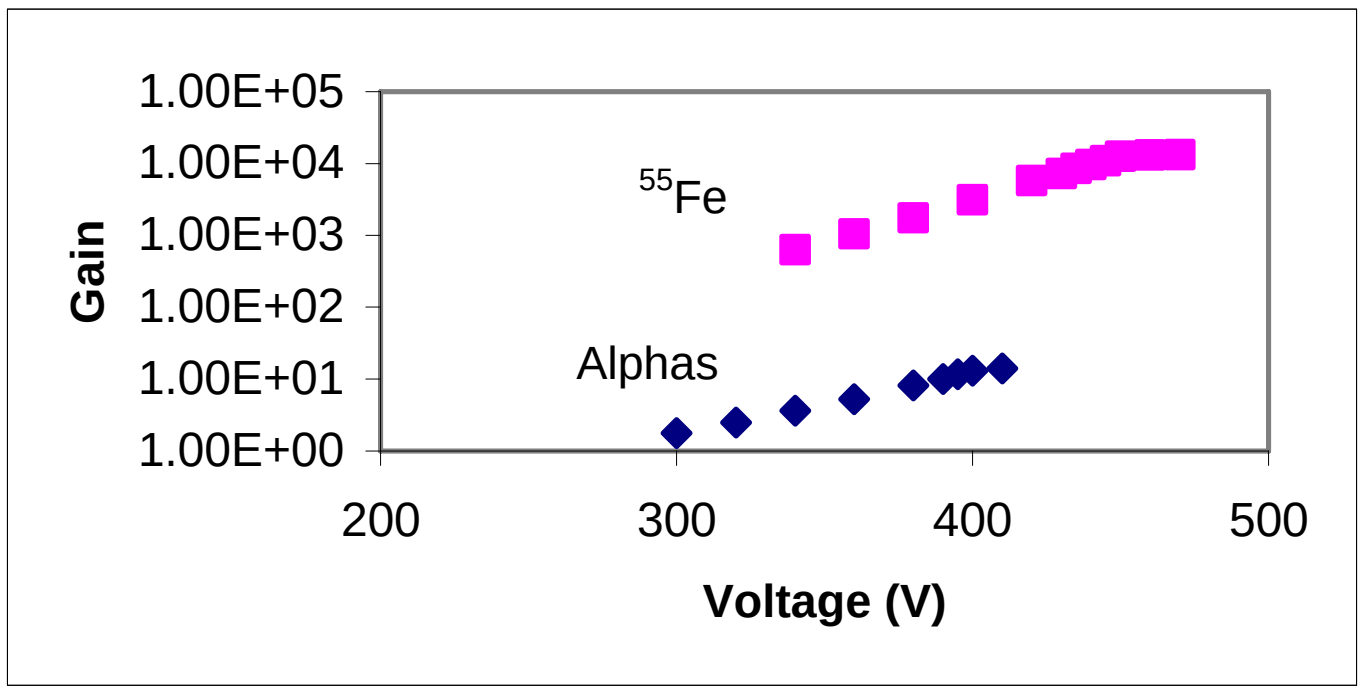

Fig. 8. Gain calculated from the measurements in Ne and the known sensitivity of a charge sensitive amplifier

One should note that during the first minute after breakdown, presumably due to the charging up effect, the gain vs. voltage may change by a factor of 3-5, but after approximately 3-5 min the gain will return to its original value. This behavior is similar to the operation of RPC chambers while in avalanche mode. This charging up effect is expected to be responsible for the overlap notice in the data in Fig. 8.

Fig. 9a shows the gain curve for double RETGEM vs. voltage applied across the RETGEM-2. (The voltage on the mesh was kept at a constant -800V, which implies a 400V potential drop across RETGEM-1.) One can see that gains approaching $10^5$ could easily be achieved with $^{55}$Fe in Ne gas. At higher gains discharges may appear; however, due to the high resistivity of the electrodes, these discharges did not harm either the detector or the preamplifiers.

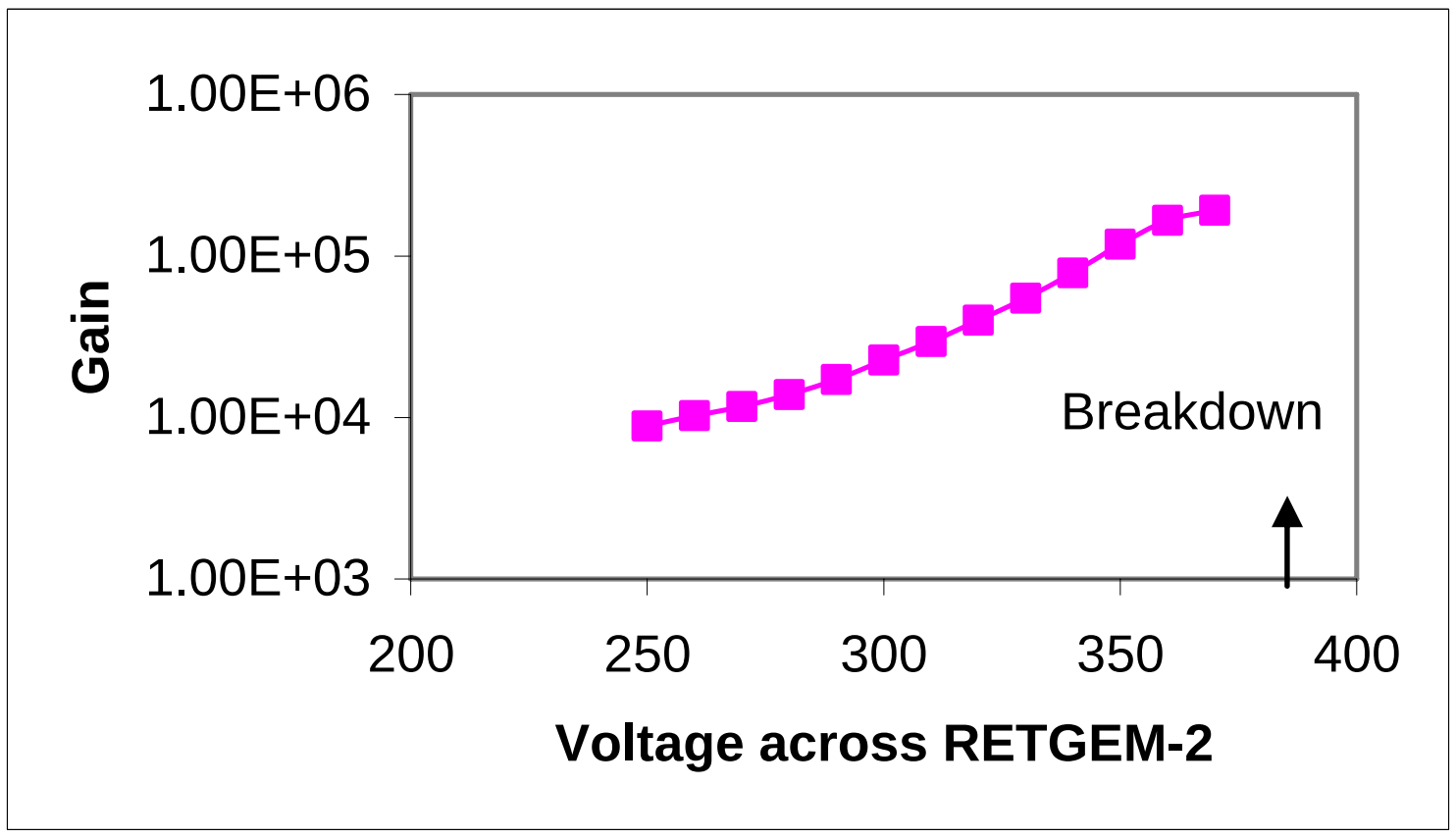

Fig. 9a. Gain curve measured with double RETGEM operating in Ne. The radioactive source was $^{55}$Fe

One might expect the that gain would be higher in the double RETGEM setup; however, the implementation of the voltage divider shown in Fig. 6 implies that the transfer potential between the bottom of RETGEM-1 and the top of RETGEM-2 was approximately 267V across an approximate separation distance of 6.3mm. By referring to previous single RETGEM data and extraction data previously accumulated using Kapton resistive GEMs of similar geometry, Fig. 9b was created to display the predicted gain given a specified extraction percentage, which is directly dependent on the transfer potential.

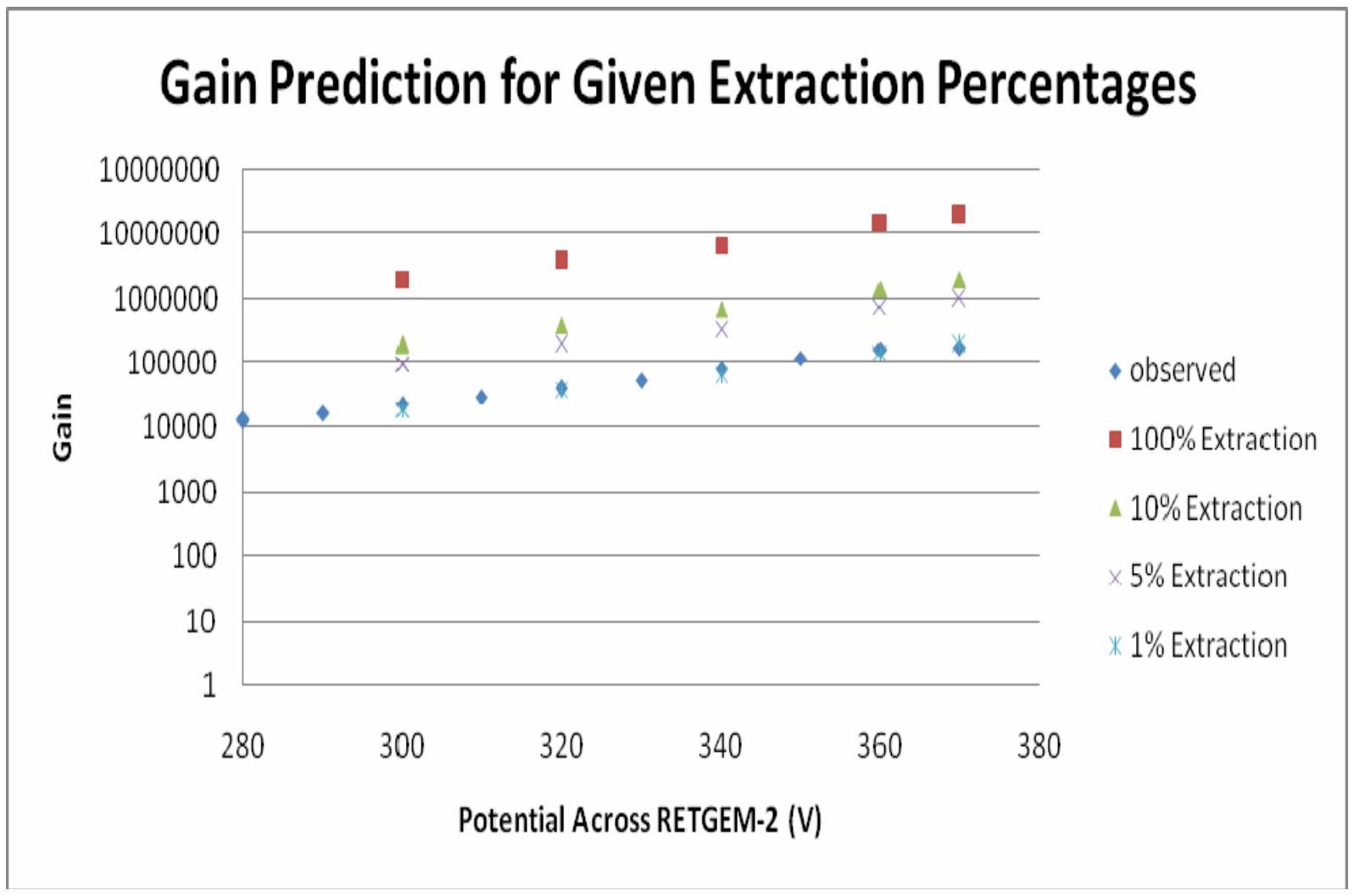

Fig. 9b. Predicted gain curves for given extraction percentages where extraction refers to the signal transfer from RETGEM-1 to RETGEM-2

Notice that the extraction percentage, or transfer percentage, for the data in Fig. 9a was approximately 1%. Assuming a modest transfer potential of 500V, which corresponds to approximately 10% signal transfer, the gain is predicted to be greater than $10^6$. However, in our measurements we prefer to keep the extraction filed at low values to avoid possible propagation of discharges from one RETGEM to another one (see [12] for more details).

Similar results were obtained in Ar and Ar+$CO_2$. Figs. 10 and 11 display the gain curves measured in a single-step RETGEM setup as function of the voltage applied across the RETGEM. One can see that gain ~$10^3$ could be achieved with $^{55}$Fe. Higher gains were achieved with a double-step RETGEM configuration. As an example, Fig. 12 and 13 show the gain curves measured with double RETGEMs operating in Ar and Ar+3%$CO_2$, respectively. One can see that gains close to$10^4$ were possible to achieve. Any discharges that appeared at higher gains were not harmful; thus, our detectors, as in the case of Kapton RETGEMs, were fully spark protected.

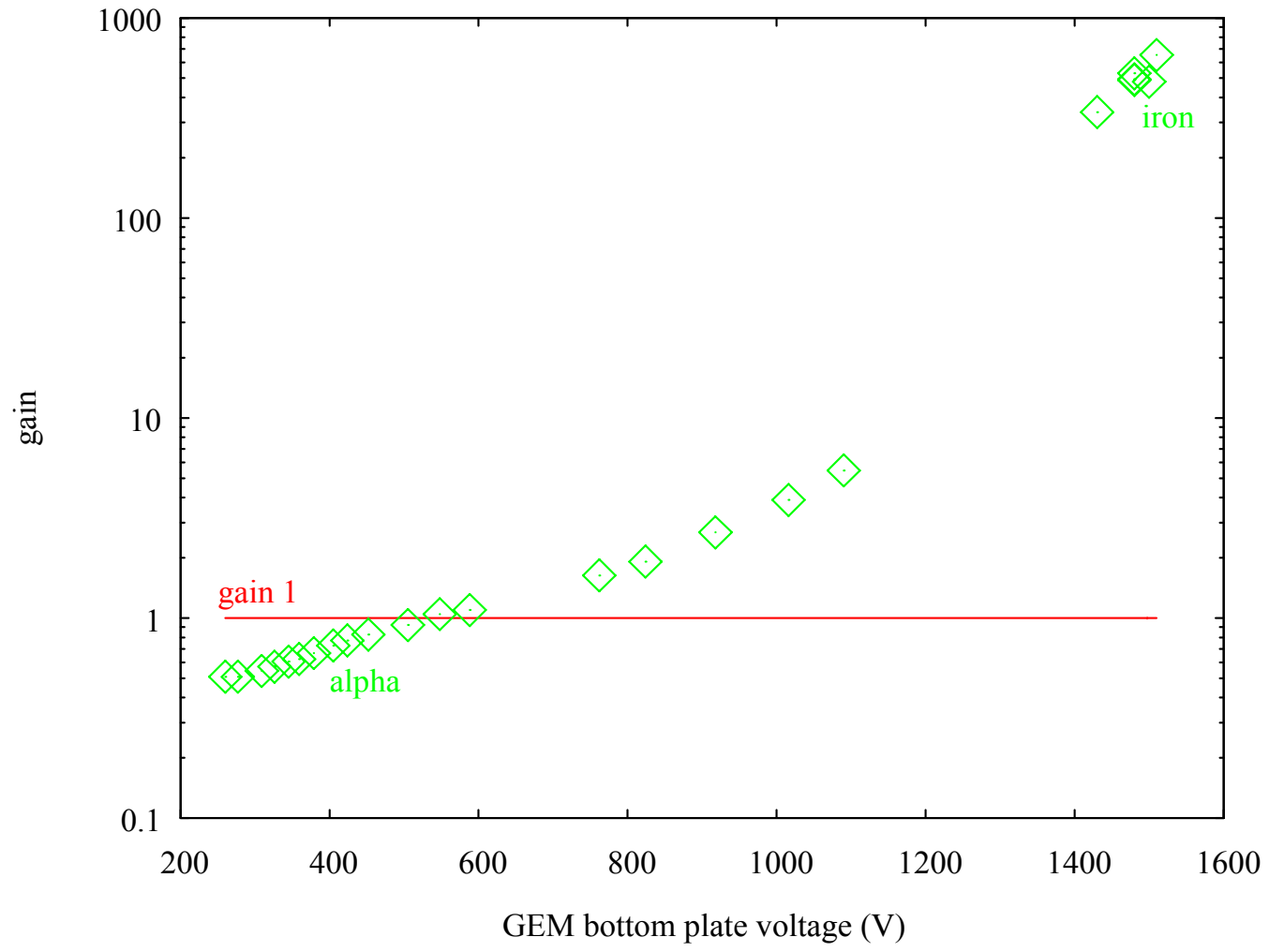

Fig.10. Gains vs. voltage measured in Ar with 5.49MeV alphas and 5.9keV x-rays

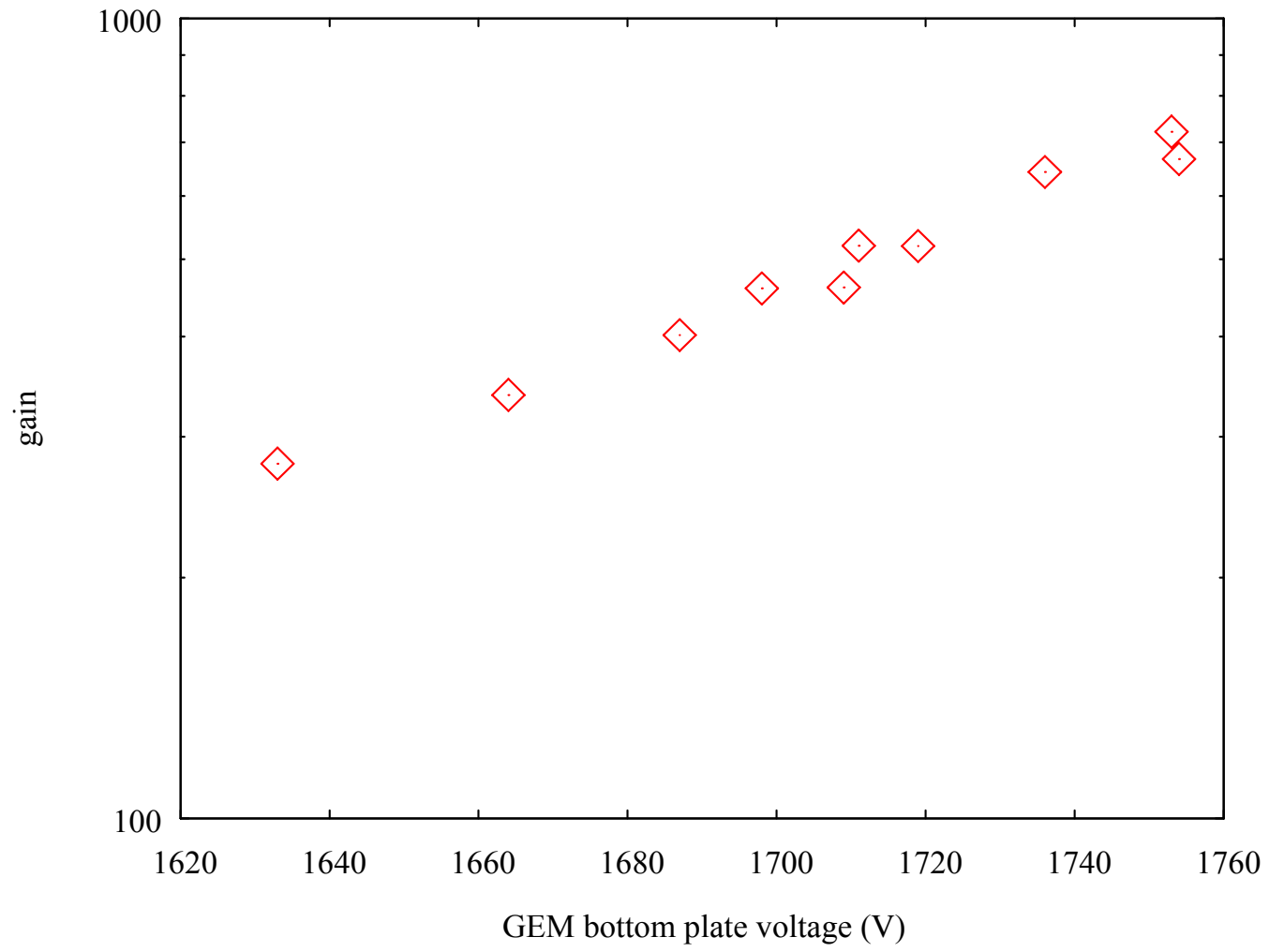

Fig.11. Gains vs. voltage measured in Ar+5%$CO_2$ with $^{55}$Fe

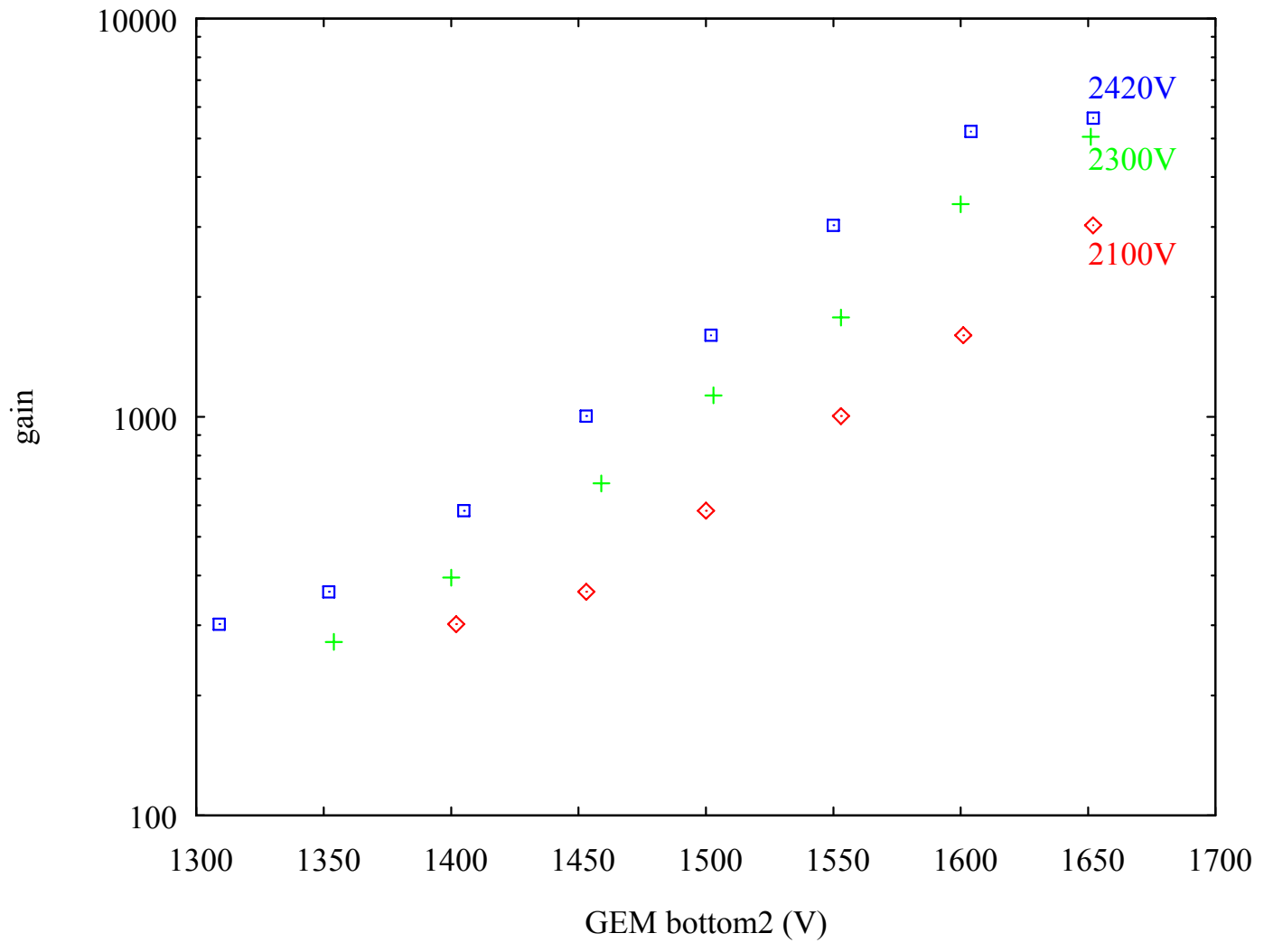

Fig 12. Gain vs. voltage curves measured in Ar with double-step RETGEM. $V_{drift}$ corresponds to the voltage applied to the voltage divider (note that $V_{drift}$ corresponds to 2420V, 2300V, or 2100V above)

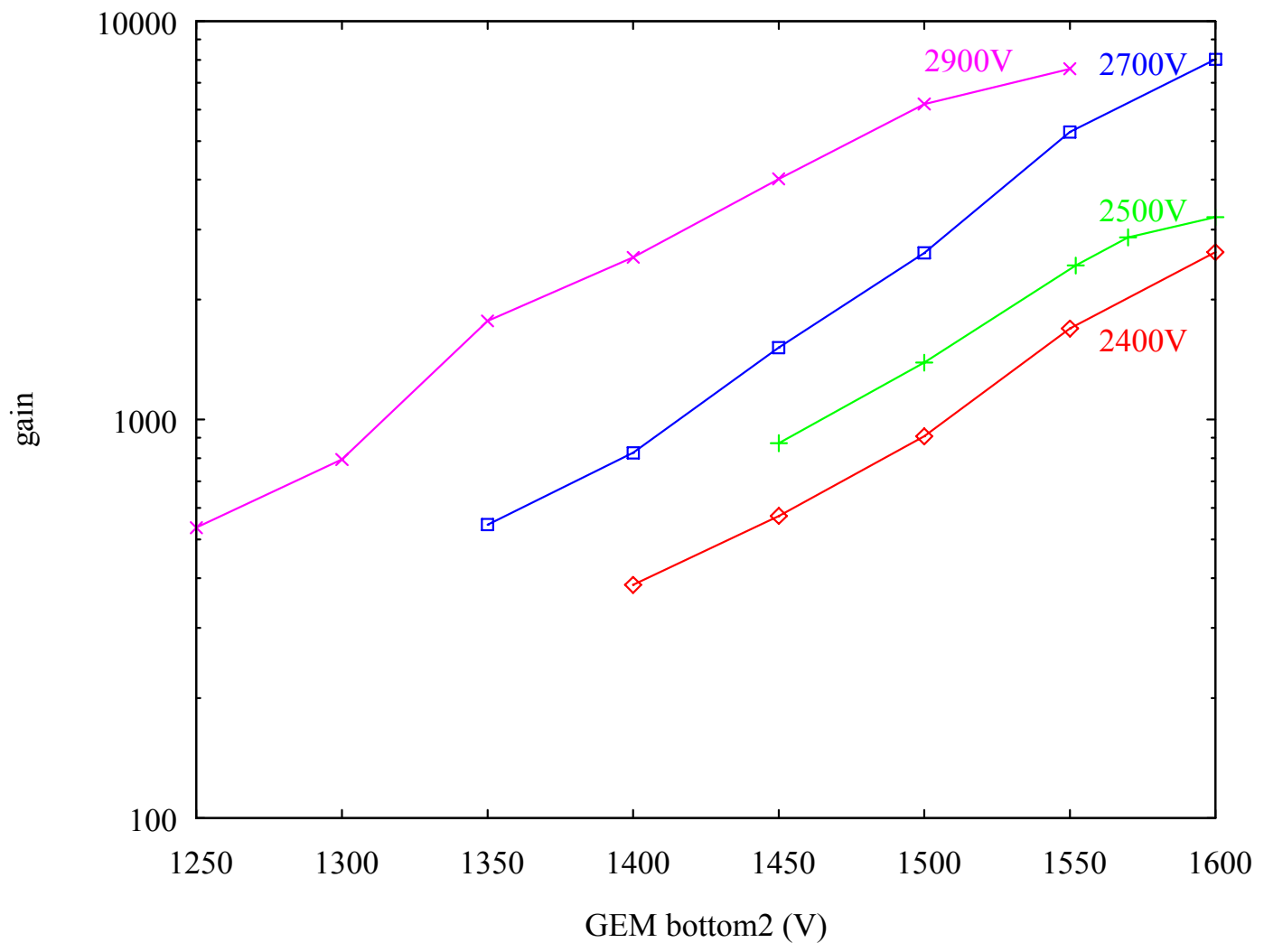

Fig.13. Gain vs. voltage curves measured in Ar+3%$CO_2$ with double step RETGEM. As in the previous figure, $V_{drift}$ corresponds to the voltage applied to the resistive chain divider

## III.2 Tests in mixtures of Ar with air and in pure air.

### *III.2a. Test with RETGEM*

In many applications and measurements it is attractive to use air-filled gaseous detectors. Examples of possible applications include dosimetry (see for example [14]) and detection of dangerous gases (see for example[15]). Recently we have discovered that TGEMs with metallic electrodes coated with oxide layers can operate in mixtures of Ar with air and even in pure air [16]. This may open many new and vital fields of practical application, including the use of RETGEMs in fire alarm systems [16].

To check if our RETGEM can operate in mixtures of Ar with air, as well as in pure air, we start measurements in pure Ar and then started diluting Ar with air. Some of our results are presented in Fig 14. This figure shows signal from the RETGEM-1 vs. time (measured at the feedthrough #3 – see Fig. 5) in pure Ar (curve #1) and then when air is gradually introduced into the chamber (curves # 2- # 12). In the case of pure Ar (curve #1), the RETGEM-1 operated at gain A=3. One can see that after the introduction of the first portion of air the signal dropped (curve #2); however, by increasing the drift voltage on feedthrough #1 and the positive voltage on the bottom of the RETGEM-1 (feedthrought #3) we were able to restore the same signal amplitude (curve #3). Thus, as an example, in this particular mixture of Ar with air the gain was A=3x3=9. We repeated this procedure 5 times more and finally the chamber was fully filled with ambient air. Each time, by increasing the voltage on the RETGEM, we restored the same signal amplitude. Thus, the gain in air is

$$A_{air}=3^{n+1}, \quad (2)$$

where n- the number of dilutions (n=6 in our case ).

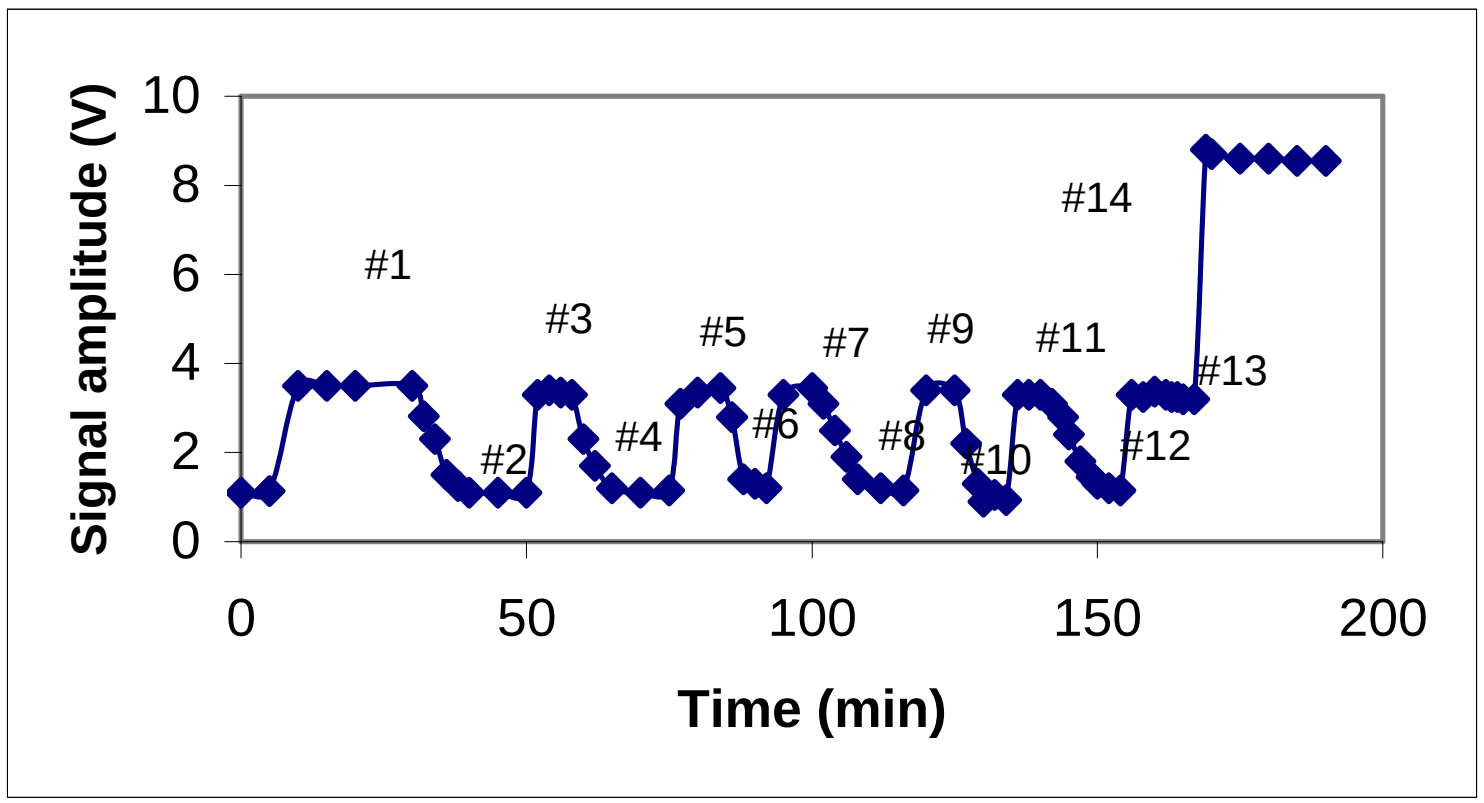

Fig. 14. Signal amplitude vs. time measured in pure Ar at gain 3 (curve#1) and then during dilution of Ar with air

Finally, in pure air we were able to additionally increase the signal by a factor of 3 (see curve #14); therefore, producing the calculated gain:

$$A_{air}=3^{n+2} \quad (3)$$
$$\text{or } A_{air}\sim 10^4.$$

Fig 15 shows a gain vs. voltage curve measured in pure air with the $^{241}$Am alpha source.

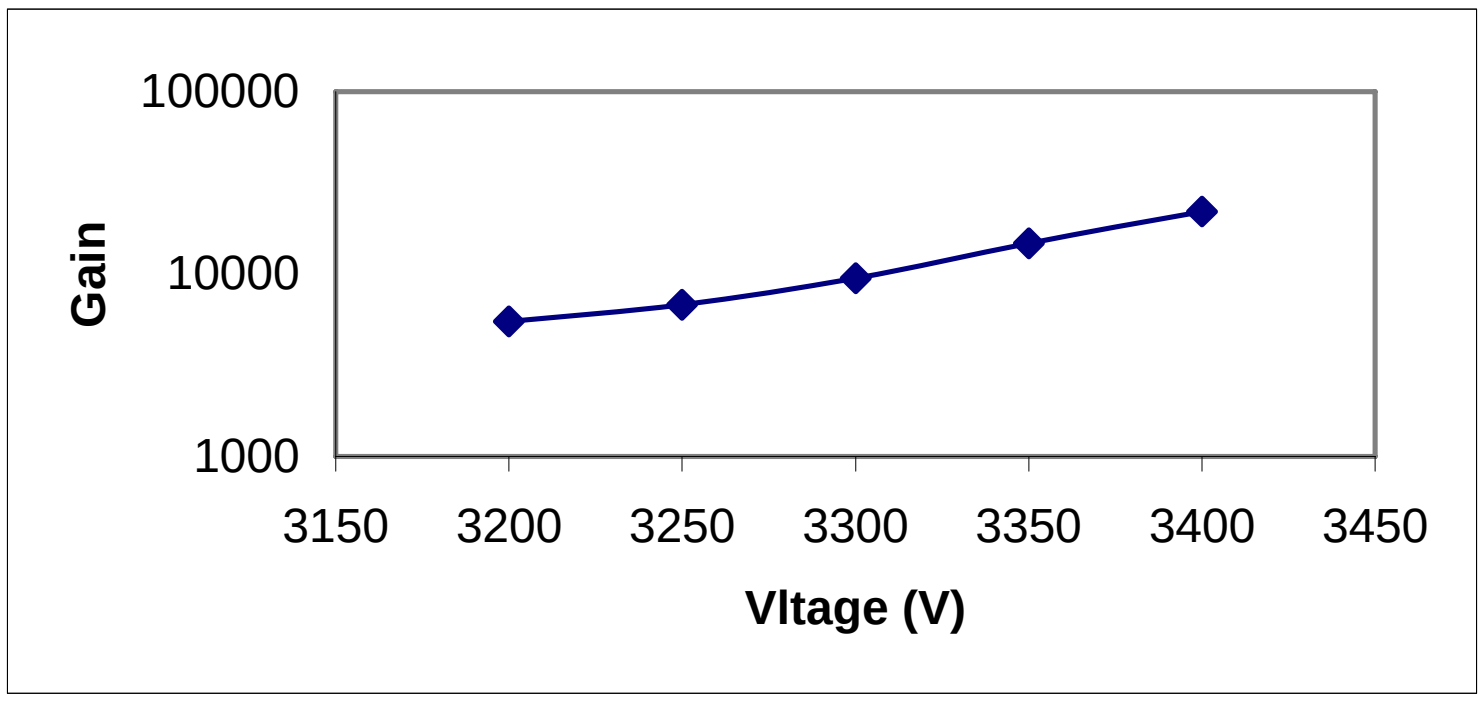

Fig. 15. Gain vs. voltage curve measured in air

It the next set of measurements we increased the drift region from 2 to 4 cm; however, it did not affect ether the maximum achievable gain or the alpha particles counting rate. Finally, we placed the $^{241}$Am source with the active part facing the drift mesh as shown in Fig. 16.

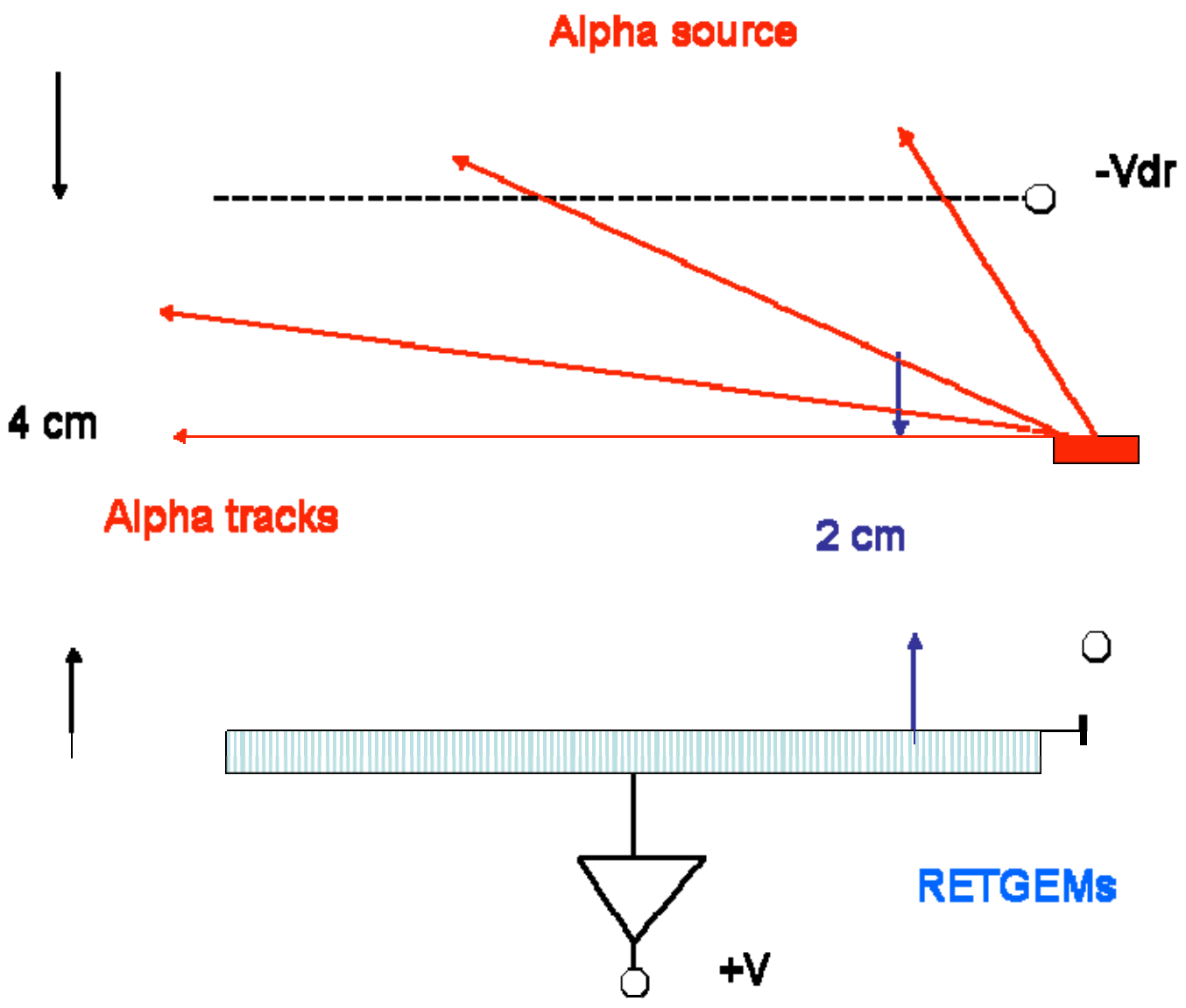

Fig 16. A schematic drawing of our set up operating in ambient air with alpha source positioned towards the drift mesh. In such geometry negative ions should drift at least 2 cm before reaching the RETGEM

In this geometry alpha particles track did not reach the RETGEM surface and negative ions should drift minimum 2 cm before reaching the holes. Again, we observed that the multiplication voltage in air was not changed and the counting rate dropped only 5-7%. These tests clearly show that negative ions formed in alpha particles track can be drifted in ambient air for a distance > 2 cm and then, when entering the RETGEM holes, trigger Townsend avalanches (detached electrons trigger the avalanches).

Note that discharges in air, as in any other tested gases, had no destructive effects and this feature may open their application in several practical devices (see conclusions).

*IV.2b. Control measurements with a metallic TGEM*

It will be interesting to understand if, in addition to the primary role of spark quenching, the resistive electrodes offer any additional roles in achieving stable high gain operation in air.

To clarify this issue, we made comparative studies with a metallic TGEM having the same geometrical characteristics as our RETGEMs. For these studies the RETGEM-1 in the test chamber was replaced by a TGEM. In the first set of measurements, a negative voltage was applied to the top electrode of the TGEM via the feedthrough #2 and a picoamperemeter was connected to the feedthrough #1. A Hg lamp was used in these measurements (see Fig. 5). Results of the measurements implementing this setup are presented as the photocurrent induced by the Hg lamp on the feedthrough #1 vs. the voltage $V_t$ applied to the top of TGEM in Fig. 17 (curve A). One can see that at $V_t$>350V the photocurrent reached the saturate value $I_t$. In the next set of measurements, the picoamperemeter was connected via the feedthrough #3 to the bottom of the TGEM. The negative voltage $V_d$=2kV was applied to the drift electrode and the current on the bottom electrode was measured as a function of the negative voltage applied to the top electrode of the TGEM. Results of such measurements are presented in Fig. 17 (curve B). One can see that at $V_t$> 2.15 kV the current from the bottom electrode begins to exceed the $I_t$ value indicating the beginning of avalanche multiplication in TGEM holes. The maximum gain we were able to reach in these conditions was $10^3$; however, by introducing a strong attenuation to the UV light intensity gains close to $10^4$ were recorded after violent sparks had appeared. In contrast to RETGEM, the operation of TGEM at voltages close to the breakdown was very unstable. Thus, one can conclude that the resistive layer not only strongly diminishes the energy of sparks, but also plays an important "stabilization" role when operating at voltages close to the breakdown level.

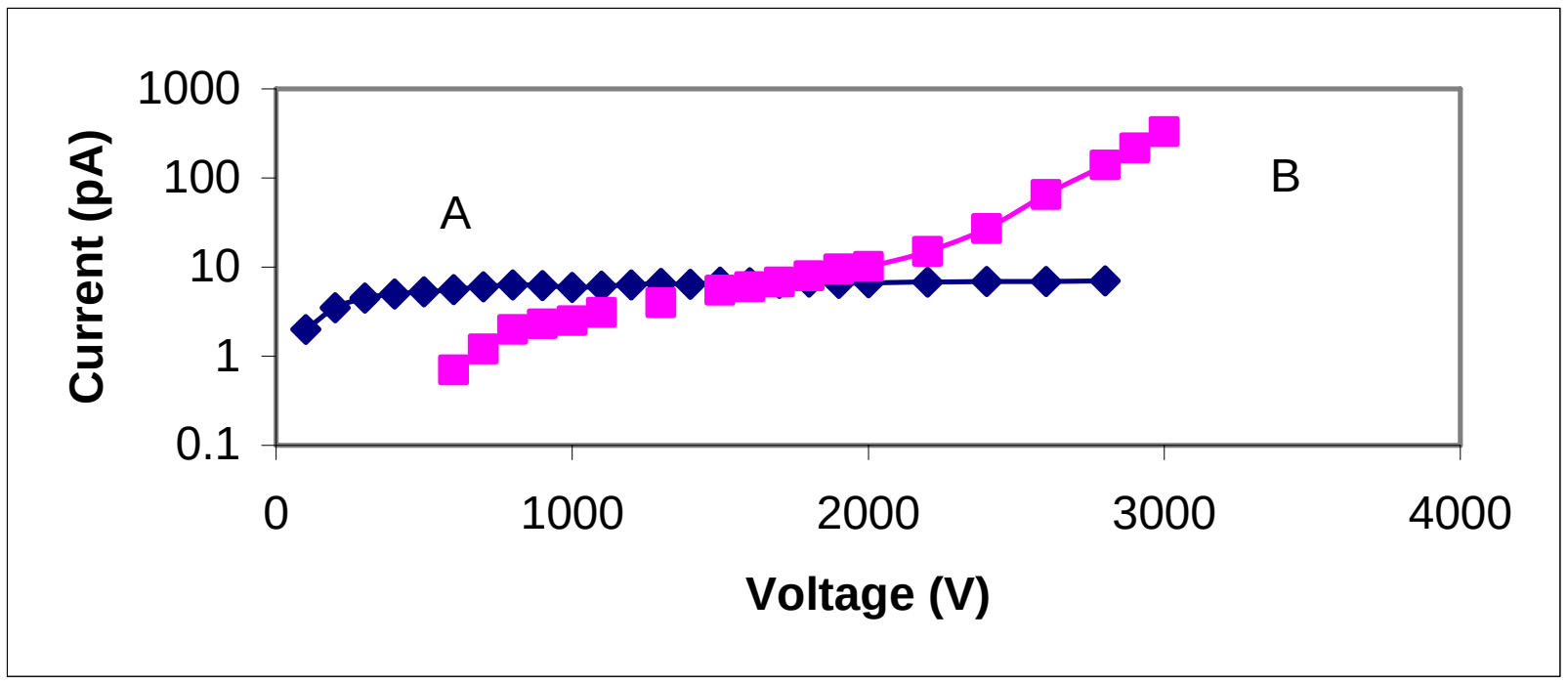

Fig. 17. Results of measurements with TGEM illuminated by a Hg lamp

## V. Conclusions

Preliminary experimental results presented in this paper haven proven that RETGEM made by a screen printing technology can operate at rather high gains even in poorly quenched gases and, most importantly, it maintains the promise of being fully spark protected. More tests are certainly needed to understand better some details in the RETGEM operation. For example, we hope to explore the stabilization role of resistive electrodes for RETGEM operating in air. Furthermore, we believe that the resistivity of 1MΩ/□ was too high, which could have attributed to the intensity of the aforementioned charging-up effect, and plan to study lower resistivity RETGEMs. Many studies are currently in preparation and will serve to promote a better understanding of the RETGEM operating characteristics. However, it is already clear that screen printing technology offers a new and desirable approach to RETGEM manufacturing by offering cost-effectiveness, convenience, and easy optimization of its resistivity and geometry. It is also important to mention that large area RETGEMs can be produced by this technology. Thus, RETGEMs made by a screen printing technique may open new avenues in both experimental and practical application. For example, its capability to operate in ambient air can be exploited in several practical devices: in sensors of dangerous gases, as a detector of alpha particles (Rn, Po) [15], or as a simple and affordable flame detector [16].